\documentclass{JINST}
\newcommand{\TEO}{$\mathrm{TeO}_2$ }

\newcommand{\qsm}{$(2310.5 \pm 1.1)$}

\title{Response of a \TEO bolometer to $\alpha$ particles}

\author{F. Bellini$^{a,b}$\thanks{Corresponding author.}, M. Biassoni$^{c,d}$, C.Bucci$^e$, N. Casali$^{a,b}$, I.Dafinei $^b$, Z.Ge$^f$, P.Gorla$^e$, F. Ferroni$^{a,b}$, F.Orio$^{a,b}$, 
C.Tomei$^b$, M.Vignati$^{a,b}$, Y.Zhu$^f$\\
\llap{$^a$}  Dipartimento di Fisica della Sapienza Universit\`{a} di Roma \\ Roma  I-00185, Italy\\
\llap{$^b$} INFN Sezione di Roma, \\ Roma  I-00185, Italy\\
\llap{$^c$}  Dipartimento di Fisica dell' Universit\`{a} di Milano-Bicocca \\ Milano  I-20126, Italy\\
\llap{$^d$}  INFN Sezione di Milano Bicocca \\ Milano  I-20126, Italy\\
\llap{$^e$} INFN Laboratori Nazionali del Gran Sasso, \\ Assergi (L'Aquila) I-67010, Italy\\
\llap{$^f$} Shanghai Institute of Ceramics Chinese Academy of Sciences, Jiading district, Shanghai 201800, P. R. China\\ 
 E-mail: \email{fabio.bellini@roma1.infn.it}}

\abstract{\TEO crystals are used as bolometers in experiments searching for Double Beta Decay without emission 
of neutrinos. One of the most important issues in this extremely delicate kind of experiments is the 
characterization of the background. The knowledge of the response to $\alpha$ particles in the
energy range where the signal is expected is therefore a must. 
In this paper we report the results on the response function of a \TEO bolometer to $\alpha$'s emitted by $^{147}$Sm dissolved in the crystal at the growth phase. A Quenching Factor of  ($1.0076\pm 0.0005$) is found, independent of the temperature in the investigated range. 
The energy resolution on $\alpha$ peaks shows a standard calorimeter energy dependence: $\sigma\; [\rm{keV}] = (0.56 \pm 0.02) \oplus (0.010 \pm 0.002)\sqrt{E[\rm{keV}]} $. Signal pulse shape shows no difference between $\alpha$ and $\beta$/$\gamma$ particles.}

\keywords{Bolometers; Quenching Factor; $\mathrm{TeO}_2$; Double Beta Decay}

\begin{document}

\section{Introduction}
\label{intro}
Bolometric detectors  \cite{Simon,Fiorini} are used in particle physics experiments to search for  rare events like Neutrinoless Double Beta Decay and Dark Matter interactions.
They are sensitive calorimeters operated at $\sim$10~mK that measure the temperature rise produced by the energy deposited in particle interactions.  

An array of bolometers made of \TEO crystals has been used in the CUORICINO experiment~\cite{Qino04,Cuoricino} to search for the Neutrinoless Double Beta Decay of $^{130}$Te. 
A precise knowledge of the response function of these bolometers to each species of particles ($\beta/\gamma$ and $\alpha$) is important for the background rejection. 
Indeed, a sizeable fraction of the background in the Double Beta Decay Region of Interest (RoI), that for  $^{130}$Te is around 2.5 MeV \cite{Scielzo,Titus}, is due to degraded $\alpha$ particles that lose part of their energy in the detector support structure and the rest in a single bolometer mimicking a signal event.

While the response of \TEO crystals to $\gamma$ interactions up to 2.6 MeV is well known from the routine calibrations performed with Th radioactive sources, there are still relevant items that should be addressed for $\alpha$ particles. The most important are the 
Quenching Factor (QF), the relative energy resolution and, finally, possible signal shape differences with respect to $\gamma/\beta$ interactions particularly in the region where the signal is expected.

The QF  is defined as the amplitude ratio between the signal produced by an $\alpha$ particle and the one produced by an electron depositing the same energy in the detector.
It is expected to be very close to unity in thermal detectors since any kind of  energy deposition should be converted into heat.

The $\alpha$'s that can be normally measured in the bolometer are either coming from decays inside the crystals or by interactions induced by external sources. Those coming from internal contaminations (usually from U and Th chains) have energies higher than 4 MeV, well above the Double Beta Decay region of interest.

The QF for $\alpha$ particles in the energy region [5.7-8.8] MeV was measured in \TEO detectors~\cite{Alessandrello1} using a $^{228}$Ra $\alpha$ radioactive source. It was found to be $1.020 \pm 0.005(stat) \pm 0.005(syst)$. No deviation from constancy of QF as a function of energy has been observed. Nevertheless this measurement is based on an extrapolation of the calibration function at energies well above the region attainable with a Th source (2.6 MeV), where non-linearities of the \TEO bolometers play an important role \cite{Vignati}. 

In this work we studied the response of \TEO bolometer to an internal source producing monochromatic $\alpha$ particles with energy close to the Q-value of  $^{130}$Te Neutrinoless Double Beta Decay. We achieved this goal dissolving into a TeO$_2$ crystal a small amount of natural samarium. The naturally occurring isotope $^{147}$Sm has an isotopic abundance of $(15.0 \pm 0.2)\%$ \cite{gov}  and an half life of $1.06 \cdot 10^{11}$y \cite{gov}. To have a reasonable $\alpha$ decay rate ($\sim$count/hour) the crystal should contain only a few micrograms of natural Sm. 

The $\alpha$ particles are emitted in the  $^{147}$Sm $\to ^{143}$Nd  +$\alpha$ transition. The Q-value of this reaction is shared between the emitted $\alpha$ particle and the recoiling nucleus. Since the decay is contained within the crystal, we measure the full transition energy of \qsm\ keV~\cite{gov,QvalueSm}. 

Beside the obvious advantage of measuring the $\alpha$ response in the actual RoI, an additional yet extremely relevant feature is present and could be fruitfully exploited. 
Bolometers, in fact, require a continuous monitoring of the operating temperature. Even tiny drifts can induce variations of the thermal gain, thus spoiling the energy resolution. To correct for this effect, a Si resistor (heater) glued to the detector surface produces heat pulses by Joule dissipation which are very similar to particle induced pulses  and the temperature drift can be corrected offline on the basis of the measured heater pulse amplitude variation~\cite{Stabilizzazione,Arn03}.  
Heaters however can experience electronic failures or grounding problems, resulting in changes of their pulse amplitude not related to temperature drifts. This calls for a frequent re-calibration of the entire system that is obviously a long time subtracted to physics data taking.

The presence of a long-lived monochromatic $\alpha$ line in the proximity of the RoI could take  the role of the heater and allows less frequent calibrations.
 
The outline of this work is the following: in Section \ref{Te02Bo}  general properties of \TEO crystals and the doping process with natural Sm  are described, in Section \ref{setup} details of the experimental setup are given  while results are summarized in Section \ref{analysis}.
 
\section{Doping of \TEO crystal with an $\alpha$-particle emitter}
\label{Te02Bo}

%\TEO crystal is well known mainly for its excellent acousto-optic (AO) properties and started to be used immediately after its first growth in 1969~\cite{Liebertz} in several AO devices such as deflectors, modulators and tunable filters for various applications. 
%Tellurium oxide is found in nature in two mineral forms, tetragonal (paratellurite) and ortho\-rhombic-dipyramidal (tellurite), from which only the first one is interesting for different applications. Paratellurite ($\alpha$-\TEO) crystal has a distorted rutile structure (asymmetric covalent Te-O bonds) with a doubling of the unit cell along the [001] direction. The positions of the tellurium are slightly shifted from the regular rutile positions, tellurium ion is fourfold coordinated by oxygen, the coordination polyhedron being a distorted trigonal bipyramid with two different bond distances. Each oxygen atom is bonded to two tellurium atoms with an angle of 140$^\circ$ between long (2.12 \AA) and short (1.88 \AA) bonds. Synthetic \TEO crystals are colorless, insoluble in water and have a density of 6.04 g/cm$^3$ which is compliant with the density calculated from measured lattice constants: a = 4.8088 \AA \ and c = 7.6038 \AA. The material is uniaxial positive with n$_{\rm  o}$ = 2.274 and n$_{\rm e}$ = 2.430, optically active and highly transparent in the range of  350 $nm$ - 5 $\mu m$ . 

Two methods are currently used for growing \TEO crystals: Czochralski and Bridgman. The growth is rather difficult, special temperature gradient conditions as well as pulling and/or rotating rates being needed in order to obtain high quality crystals. Further post-growth thermal treatments are applied to the as grown ingots aimed at quenching the tendency of \TEO single crystals to cracking caused by high anisotropy of thermal coefficients. Inhomogeneities in the crystal due to the incorporation of Pt dissolved from the crucible is another problem frequently reported~\cite{Foldvari1} and careful control of convection currents in the melt is needed to avoid the incorporation of the gas bubbles to the crystals, especially in the case of Czochralski growth.

%In the field of astroparticle physics, \TEO crystals possess the very good thermal and mechanical properties requested by the sophisticated cryogenic setup used in underground experiments searching for rare events. The high natural abundance of the $\beta\beta$ emitting isotope $^{130}$Te (33.8\%) \cite{gov} eliminates the need for complicated and costly enrichment processes when using \TEO crystals as calorimeters for the search of Neutrinoless Double Beta Decay.  

\TEO crystals do not easily permit impurities inclusion in the lattice. The paratellurite structure, especially the asymmetric covalent Te-O bonds, limits the incorporation of the dopants to extremely low levels which results in very low segregation coefficients (k$\sim$10$^{-2}$). To avoid precipitation or aggregation of foreign ions during crystal growth only small amounts of dopant can be added to the melt and the resulting built-in concentration is often below the limit for chemical analysis~\cite{Foldvari1}. Only a few successful doping of \TEO are reported in literature for ions like Fe or Cr~\cite{Foldvari2} and Mg, Mn, Nb, Zr~\cite{Dafinei}.
As a general rule for selecting the $\alpha$ dopant, two criteria should be taken into consideration: radioactive properties of the nuclide and the incorporation to the host lattice of the ion.

In our particular case, where the detector has to be applied to the search for very rare events, supplementary constraints are imposed on the radio-purity of the chemical compound used for the doping. Moreover, for cryogenic applications of \TEO crystals, the incorporation of uneven ions in the \TEO lattice is to be avoided because of the possible problems of cooling paramagnetic materials at very low temperatures.

\begin{figure}[htdp]
\centering
\includegraphics[scale=.4]{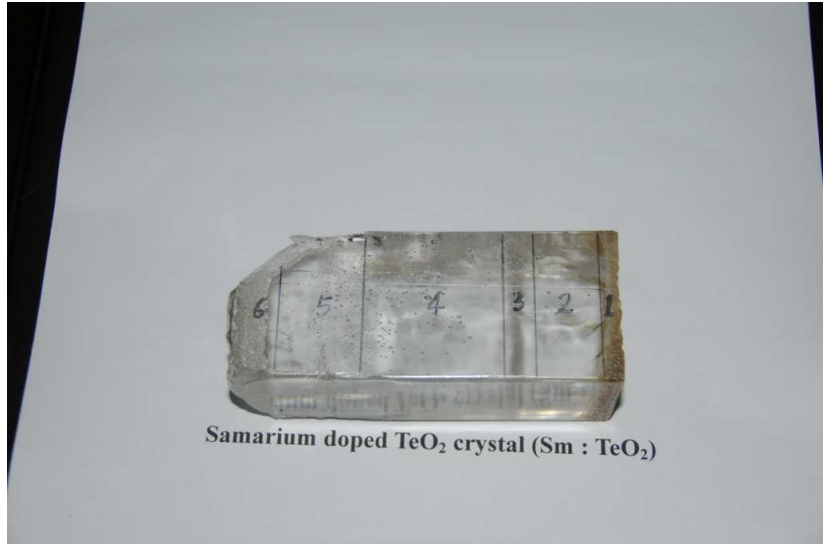}
\caption{\small{Sm doped \TEO crystal as grown. Different samples were cut for analysis. The cloudy (no. 3) region is due to a thermal instability during the growth process.}}\label{Fig1}
\end{figure}

The Sm doped crystal reported in the present work was grown by a modified Bridgman method described in detail in~\cite{Jiayue}. 
For the growth, 6N purity \TEO powder was used. Tellurium oxide raw material was synthesized at SICCAS, Shanghai China, in a process described in detail in~\cite{Yaoqing}. 
The crystal was grown in a dedicated furnace used for R\&D purpose where ingots of typically 30 x 25 x 120~mm can be obtained. The crystal growth followed the protocol described in 
detail in~\cite{Arnaboldi}. Sm dopant was added to the raw material powder in oxide form using Sm$_2$O$_3$ powder of 6N purity. The doping process was performed in two steps. 
The Sm$_2$O$_3$/\TEO powder was first melted in order to obtain an ingot of homogeneous oxides mixture from which a small sample was taken and used as dopant for the final crystal 
growth with a nominal dopant concentration of 5$\cdot$10$^{-6}$~g/g of Sm$_2$O$_3$/\TEO. A charge of approximately 400~g (crystal seed excluded) was used for the final growth which
 resulted in an ingot of approximately 65~cm$^3$. The grown ingot was colorless and free of cracks, bubbles and/or inclusions except for a small region due to a thermal instability 
 during the growth process (see Fig.~\ref{Fig1}). It was subjected to special post-growth thermal treatment in order to avoid cracks during mechanical processing. 
 A sample of approximately 30x24x28~mm$^3$ (m=116.65~g) was extracted for cryogenic measurement from the region no.4 of the as grown crystal. 
 The sample was cut and X-ray oriented with a precision better than 0.5$^\circ$ following the procedure described in~\cite{Arnaboldi} except for the chemical etching. 
 Different samples were taken from each region (1 to 6) for ICP-MS measurements aimed at checking the general quality of the crystal and especially the uniform 
 distribution of dopant along the growth direction. ICP-MS measurements were made using an "Agilent Technologies 7500 Series" instrument \cite{Arnaboldi}.  The measurements were made on samarium isotopes free of instrumental interferences: $^{149}$Sm, $^{152}$Sm and $^{154}$Sm. For each isotope  the Sm concentration was inferred assuming the natural abundance. The mean value is found to be of the order of 30~ppb which gives for the 116.65~g sample approximately 3.5 $\mu$g Sm.

\section{Experimental set-up and Data Acquisition}
\label{setup}
The Sm doped \TEO crystal was operated in a cryogenic setup located deep underground in the Hall A of National Laboratory of INFN at Gran Sasso. 

The crystal was equipped with two Neutron Transmutation Doped Ge thermistors (NTD)~\cite{NTD} of 3$\times$3$\times$1~mm$^3$,  thermally coupled to the crystal surface with 9 epoxy glue spots ($\sim$0.8~mm diameter). The electrical conductivity of NTD, which is due to variable range hopping (VRH) \cite{VRH} of the electrons, depends strongly on the temperature. The resistance changes with temperature according to \cite{Itoh}
\begin{equation}
\hspace{1cm}    R=R_{\rm 0} \cdot   e^ { \left( \frac{T_{\rm 0}}{T} \right)^{1/2}}
\end{equation}
and then resistance variation can be used to measure effectively the heat signal produced by interacting particles.
The $R_{\rm 0}$ and $T_{\rm 0}$ of the two NTDs were measured to be respectively $\sim$1$\Omega$ and $\sim$3K. At the working temperature of $\sim$10 mK the value of R is $\sim$ 100 M$\Omega$. 

A resistor of  $\sim$ 300~k$\Omega$, realized with an heavily doped meander on a 2.33$\times$2.4$\times$0.6$\times$~mm$^3$ silicon chip, was glued to the crystal and used as a heater to stabilize the gain of the bolometer.

The crystal was mounted in an Oxygen Free High Conductivity (OFHC) copper structure and kept in position by PTFE tips. The L shape of the PTFE pieces was chosen to profit from the high thermal contraction of the Teflon keeping tightly the crystal. 

Finally, the detector was mounted and operated at $\sim$ 10~mK, cooled by an Oxford 1000 $^{3}$He/$^{4}$He dilution refrigerator. The crystal holder was mechanically decoupled from the cryostat in order to minimize noise vibrations induced by the cryogenic facility. A weak thermal coupling between the mixing chamber of the dilution refrigerator and the crystal holder was realized by means of thin high conductivity copper strips. The holder was also equipped with one NTD thermometer and one heater. In this way it was possible to stabilize the temperature of the crystal holder using a feedback device~\cite{Holderstab}. 

The cryostat was heavily shielded both internally and externally in order to decrease the $\gamma$ background on the detector coming from radioactive materials.
Below and above the detector there was a $\sim$ 10~cm thick layer of low-activity ancient Roman lead~\cite{Pbromano1,Pbromano2}.  Around the sides of the detector holder an additional 1.2~cm thick cylindrical Roman lead shield was present. The cryostat was surrounded by an additional external shielding composed by  10~cm low-activity lead, 10~cm standard lead followed by 10~cm borated polyethylene. The latter allows to reduce the neutron flux on the detector thanks to the high efficiency in thermalizing fast neutrons and to the high neutron capture cross section for thermal neutrons of $^{10}$B.
The entire setup was enclosed in a Faraday cage to reduce electromagnetic interference. 

The thermistors were biased through two room temperature 27~G$\Omega$ load resistors. The large ratio between their resistances and those of the thermistors allows to have negligible parallel noise.

The read-out of the thermistors was performed through a room temperature DC coupled differential front-end~\cite{Programmable} followed by a second stage of amplification both located on the top of the cryostat. After the second stage, and close to the DAQ (a 18 bit NI-PXI 6284 ADC unit), a 6 pole roll-off active Bessel filter acted as antialiasing filter. 

The entire waveform of each triggered pulse was sampled with a rate of  1 kHz and recorded. The typical bandwidth is approximately 10~Hz, with signal rise and decay times of order of 40 and 200~ms, respectively.

The two thermistors are read by independent electronic channels and will be denoted in the following as Channel 1 and Channel 2.

A more detailed description of the electronics and the cryogenic facility can be found in Ref.~\cite{Qino04,Cuoricino}.

\section{Data Analysis and Results}
\label{analysis}
Two separate sets of data have been collected at different holder temperatures (run1: $\sim$10~mK, run 2: $\sim$15~mK ) in order to investigate possible temperature 
dependencies of the Quenching Factor and differences of the pulse shape between $\alpha$ and $\beta/\gamma$ particles.

\begin{figure}[htdp]
\centering
\includegraphics[scale=.6]{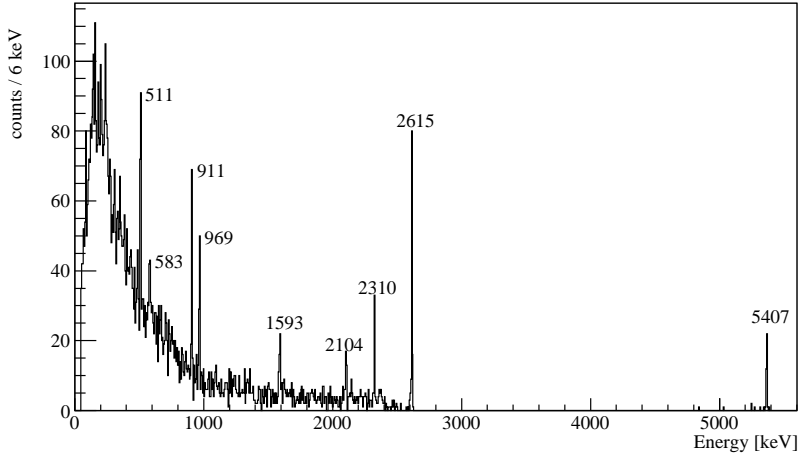}
\caption{Calibration spectrum of Channel 1 - run 1. The calibration is performed with external  $^{232}$Th $\gamma$ sources. The main peaks used for calibration are labeled together with the $^{147}$Sm $\alpha$ line at 2310~keV and the $^{210}$Po $\alpha$ line at 5407~keV.}
\label{Fig2}
\end{figure}

The energy calibration is performed using four $^{232}$Th $\gamma$ sources inserted inside the external lead cryostat shield. 
The pulse amplitude (A) is estimated by means of an Optimum Filter technique~\cite{Gatti}. The Channel 1 calibration spectrum in run 1 is shown in  Fig.~\ref{Fig2}. Gamma lines from the $^{232}$Th decay chain are visible in the spectrum and listed in Table~\ref{Tab1}. The $^{147}$Sm $\alpha$ line is also clearly visible as well as another $\alpha$ line at 5.407~MeV from $^{210}$Po. This contamination originates during the crystal growth. The line from $^{210}$Po is not considered in our QF calculation since it lies in an energy region where the calibration is not reliable at the desired level of accuracy. 

The calibration peaks are fitted using a Gaussian function + linear background.

\begin{table}[htdp]
\begin{center}
\begin{tabular}{|c|c|}
\hline
Isotope - source & Energy [keV] \\
\hline
\hline
e$^{+}$ e$^{-}$ annihilation & 511.0 \\
\hline
$^{208}$Tl & 583.2 \\
\hline
$^{228}$Ac & 911.2 \\
\hline
$^{228}$Ac & 968.9 \\
\hline
2614.5 keV DE & 1592.5 \\
\hline
2614.5 keV SE & 2103.5 \\
\hline
$^{208}$Tl & 2614.5 \\
\hline
\end{tabular}
\caption{\small{Calibration peaks from the $^{232}$Th source.}}\label{Tab1}
\end{center}
\end{table} 

The calibration function is a third order polynomial with zero intercept. The Quenching Factor is extracted from a simultaneous fit to the $\gamma$'s  and the $\alpha$ line adding the QF as a free additional parameter. To properly take into account uncertainties on fitted amplitudes (A$_i$) and  transition energies (E$_i$), the following likelihood function has been minimized:
\begin{equation}
-\log\mathcal{L} = -\sum_i \log\frac{1}{\sqrt{2 \pi} \sigma_i} e^{-\left(\frac{E_i -P(A_i)}{\sqrt{2}\sigma_i}\right)^2} -\log\frac{1}{\sqrt{2 \pi} \sigma_{\alpha}} e^{-\left(\frac{E_{\alpha} -P(A_{\alpha})/QF}{\sqrt{2}\sigma_{\alpha}}\right)^2}
\end{equation} 
where the index runs over the calibration peaks, P(A) is the polynomial calibration function to be estimated and $\sigma_{i,\alpha}$ is computed as:

\begin{equation}
\sigma_{i,\alpha} =  \sqrt{\sigma_{E_{i,\alpha}}^2 + \left( \frac{\partial P}{\partial A}  \sigma_{A_{i,\alpha}} \right)^2}
\end{equation} 
 
The QF calculated for both channels and runs is reported in Table~\ref{Tab2}. 
The correlation between the QFs has been evaluated on the energy distribution of coincidence events  in the $\alpha$ peak, recorded simultaneously by the two channels.
Taking into account the full covariance matrix, a  ${\rm QF } =  1.00755\pm 0.00066$  and  QF = $ 1.00765\pm 0.00066$ is found for  run1 and run2 respectively. The error is dominated by the uncertainty on the $^{147}$Sm Q-value. 
No temperature dependence is observed; the average QF is $ 1.0076\pm 0.0005$.

\begin{table}[htdp]
\begin{center}
\begin{tabular}{|c|c|c|c|}
\hline
Run & Channel & QF & QF error\\
\hline
\hline
1 & 1 & 1.0076 & 0.0007 \\
\hline
1 & 2 & 1.0075 & 0.0007 \\
\hline
2 & 1 & 1.0078 & 0.0007 \\
\hline
2 & 2 & 1.0075 & 0.0007 \\
\hline
\end{tabular}
\caption{\small{QF calculated for the two channels for both runs. The error is dominated by the uncertainty on the $^{147}$Sm Q-value (see Introduction).}}\label{Tab2}
\end{center}
\end{table}

\subsection{Energy Resolution and Pulse Shape}
We have performed a study on the dependence of the energy resolution of the $\gamma$ ($^{232}$Th) and the $\alpha$ ($^{147}$Sm and $^{210}$Po) peaks as a function of energy.

The Optimum Filter~\cite{Gatti}, used for amplitude estimation, reduces the noise superimposed to the signal, maximizing the signal to noise ratio. The ultimate resolution is governed by the noise at the filter output, and does not depend on the pulse amplitude.
We estimated this resolution, $\sigma_{OF}$, on data samples recorded  randomly and without triggered events.

\begin{figure}[htdp]
\centering
\includegraphics[scale=.4]{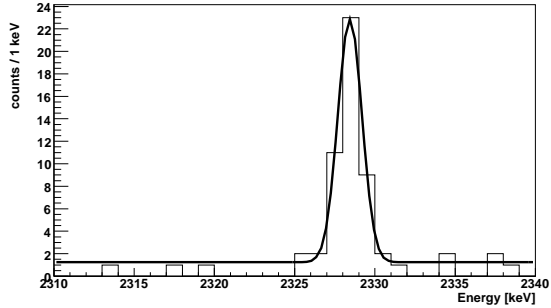}
\caption{\small{Gaussian + linear background fit of the $^{147}$Sm line for Channel 1 - run 2. The energy (not corrected for the Quenching Factor) is {$(2328.4\pm 0.14)$} keV, with resolution of {$(0.75\pm 0.12)$} keV.}}\label{Fig3}
\end{figure}

\begin{figure}[htdp]
\centering
\includegraphics[scale=.4]{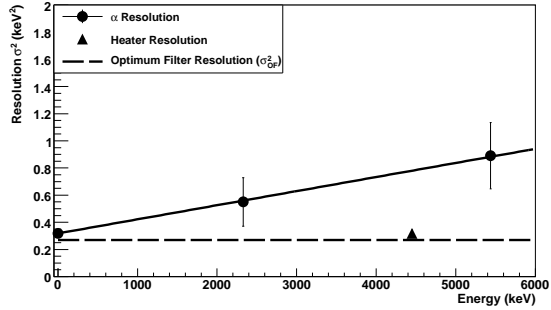}
\caption{\small{Squared energy resolution ($\sigma^2$) for $\alpha$'s (dots) and heater (triangle) as a function of energy for Channel 1 - run 2. The theoretical Optimum Filter resolution ($\sigma^2_{OF}$) is shown for comparison (dashed line). The linear fit on $\alpha$ points is performed assuming that the resolution at zero energy is equal to the heater resolution ($\sigma^2_H$) with the corresponding error.}}\label{Fig4}
\end{figure}

The energy resolution of  $\alpha$ and heater peaks has been calculated with Gaussian + linear background fits (see for example the fit on the $^{147}$Sm line in Fig.~\ref{Fig3}). The quantity $\sigma^2$ is shown as a function of energy in Fig.~\ref{Fig4}.
The squared heater resolution $\sigma^2_H$ (black triangle)  is well consistent with $\sigma^2_{OF}$ (dashed line), indicating that no physical process other than the noise
contributes to it. On the other end, the resolution of $\alpha$ lines lies above $\sigma^2_{OF}$  and exhibits a dependence on the energy. 

A linear fit on the $\alpha$  points of Fig.~\ref{Fig4} is performed, assuming that the resolution at zero energy is equal to the heater resolution ($\sigma^2_H$) with the corresponding error. 
The energy dependence of the $\alpha$ resolution  is well in agreement with the behavior of a calorimeter, where not all the energy is eventually collected: 

$$
\sigma\; [\rm{keV}] = (0.56 \pm 0.02) \oplus (0.010 \pm 0.002)\sqrt{E[\rm{keV}]} .
$$
This behaviour indicates that in a macrobolometer, where all the energy is expected to be transformed into heat and measured, some phenomena are responsible for a missing part.
We performed the same analysis on Channel 2 and found a consistent energy dependence: ~$(0.012 \pm 0.003) \sqrt{\rm{keV}}$.  

A more intriguing feature emerges from the energy resolution measured on the $^{208}$Tl $\gamma$ line. Its value is worse than the one of an $\alpha$ line of equivalent energy:
\begin{eqnarray*}
\sigma^2_{Tl} &=& (5.9 \pm 2.3)\;\rm{keV}^2 \\
\sigma^2_{\alpha (2614.5\;\rm{keV})} &=& (0.6 \pm 0.1)\;\rm{keV}^2 
\end{eqnarray*}
The crystal is however too small and the statistics too limited to allow a detailed study of the energy resolutions of the $\beta$/$\gamma$ branch.

We investigated the difference in  pulse shape between $\alpha$ and $\beta$/$\gamma$ particles.
In bolometric detectors  the shape of the signal is not constant with energy  as detailed in \cite{Vignati}. 
Shape parameters, therefore,  have been compared  for events in the $^{147}$Sm $\alpha$ peak ([2320-2334] keV) and in  the sideband regions ([2200-2320] keV, [2340-2460] keV), mainly populated by $\beta$/$\gamma$'s.

The rise and decay times, computed as the time difference between the 10\% and the 90\% of the leading edge and the 90\% and 30\% of the trailing edge respectively, are reported in  Table~\ref{Tab3}. The error is dominated by the resolution of the sampling period. No difference is found within the error.

\begin{table}[htdp]
\begin{center}
\begin{tabular}{|c|c|c|}
\hline
Particle  & Rise Time [ms] & Decay Time [ms] \\
\hline
\hline
$\alpha$  & 49.0 $\pm$ 0.4  & 219.6 $\pm$ 0.4 \\
\hline
$\beta$/$\gamma$  & 49.0 $\pm$ 0.4& 219.9 $\pm$ 0.4\\
\hline
\end{tabular}
\caption{\small{The rise and decay times for $\alpha$  and $\beta$/$\gamma$ particles in the  $^{147}$Sm $\alpha$ region and in the sideband regions. }}\label{Tab3}
\end{center}
\end{table}

\section{Conclusions}

We characterized the response function of a \TEO bolometer to $\alpha$ particles with energy close to the Q-value of the $^{130}$Te Neutrinoless Double Beta Decay.
To achieve this goal  we dissolved  into a TeO$_2$ crystal  a small amount of $^{\rm nat}$Sm at the growth phase.

A Quenching Factor for $\alpha$ particles of  ($1.0076\pm 0.0005$) is found, independent of the temperature in the investigated range. 

The energy resolution  on  $\alpha$ peaks  shows a standard calorimeter energy dependence: $\sigma\; [\rm{keV}] = (0.56 \pm 0.02) \oplus (0.010 \pm 0.002)\sqrt{E[\rm{keV}]} $.
Resolution on a $\gamma$ peak is found to be worse compared to an $\alpha$ line of equivalent energy, but the crystal is too small and the statistics too limited to draw any conclusion.
Signal pulse shape shows no difference between $\alpha$ and $\beta$/$\gamma$ particles.

Finally, the presence of a long-lived monochromatic $\alpha$ line in the proximity of the $^{130}$Te Double Beta Decay RoI, could be used to correct thermal gain variation thus allowing  less frequent calibrations.

%\acknowledgments

\end{document}